\documentclass[cameraready]{Interspeech}

\usepackage{booktabs}
\usepackage{multirow}
\usepackage{amsmath}
\usepackage{amssymb}
\usepackage{pifont}
\usepackage{graphicx}
\newcommand{\cmark}{\ding{51}}

\title{StreamVoiceAnon+: Emotion-Preserving Streaming Speaker Anonymization
via Frame-Level Acoustic Distillation}

\author[affiliation={1,2}]{Nikita}{Kuzmin}
\author[affiliation=3]{Kong Aik}{Lee}
\author[affiliation=1]{Eng Siong}{Chng}

\address{$^1$Nanyang Technological University, Singapore \\
$^2$Institute for Infocomm Research (I2R), A*STAR, Singapore \\
$^3$The Hong Kong Polytechnic University, Hong Kong}
\email{s220028@e.ntu.edu.sg}


\makeatletter
\def\bstctlcite#1{\@bsphack
  \@for\@citeb:=#1\do{%
    \edef\@citeb{\expandafter\@firstofone\@citeb}%
    \if@filesw\immediate\write\@auxout{\string\citation{\@citeb}}\fi}%
  \@esphack}
\makeatother

\begin{document}
\bstctlcite{IEEEexample:BSTcontrol}

\maketitle

\begin{abstract}
We address the challenge of preserving emotional content
in streaming speaker anonymization (SA).
Neural audio codec language models trained for audio continuation
tend to degrade source emotion:
content tokens discard emotional information,
and the model defaults to dominant acoustic patterns
rather than preserving paralinguistic attributes.
We propose supervised finetuning with neutral-emotion utterance pairs
from the same speaker,
combined with frame-level emotion distillation
on acoustic token hidden states.
All modifications are confined to finetuning,
which takes less than 2 hours on 4 GPUs
and adds zero inference latency overhead, while
maintaining a competitive 180ms streaming latency.
On the VoicePrivacy 2024 protocol,
our approach achieves a 49.2\% UAR (emotion preservation)
with 5.77\% WER (intelligibility),
a +24\% relative UAR improvement over the baseline
(39.7\%$\to$49.2\%) and +10\% over the emotion-prompt variant (44.6\% UAR),
while maintaining strong privacy (EER 49.0\%).\footnote{Demo and code: \url{https://anonymous3842031239.github.io}}
\end{abstract}

\section{Introduction}

Speaker anonymization (SA) aims to transform an input speech (source) to conceal speaker identity
while preserving linguistic content and paralinguistic attributes
such as emotion~\cite{tomashenko2022vpc2020,panariello2024voiceprivacy,tomashenko2026vpc3}.
Real-time SA is essential for privacy-preserving applications
including teleconferencing, call centers, voice assistants,
and online mental health counseling~\cite{meyer2025usecases}, where low latency is critical.
While privacy and intelligibility have been the primary focus of SA research,
emotion preservation is equally important for natural communication.
Losing emotional cues can significantly degrade user experience
and communication effectiveness in real-world deployments~\cite{cai2024privacy}.

Recent advances in neural audio codec (NAC) language models
have enabled streaming SA with competitive privacy-intelligibility
trade-offs~\cite{panariello2024nac,kuzmin2026streamvoiceanonenhancingutilityrealtime},
unlike offline approaches that require full utterance
context~\cite{yao2025easy,yao2024npuntu}.
These models process speech as interleaved content
and acoustic token sequences,
allowing autoregressive (AR) generation with controllable speaker identity.
However, two fundamental limitations emerge for emotion preservation:
(1) the audio continuation training paradigm encourages the model
to degrade emotion preservation for the source utterance, and
(2) the VQ bottleneck in neural codecs discards fine-grained
acoustic details that carry emotional information~\cite{zhang2024speechtokenizer,gaznepoglu2025why}.

To mitigate this limitation, prior work \cite{kuzmin2026streamvoiceanonenhancingutilityrealtime} employed multiple emotion-diverse prompts
at inference, which partially improves emotion preservation
but significantly degrades intelligibility
and requires emotion-labeled prompts that are harder to obtain
compared to diverse neutral prompts commonly used for anonymization.

\begin{figure}[t]
    \centering
    \includegraphics[width=\linewidth]{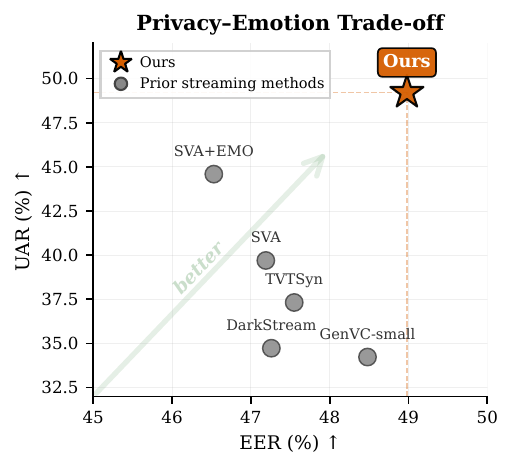}
    \caption{Privacy-emotion trade-off for streaming speaker
    anonymization methods.
    Our method (orange star) compared to prior streaming methods
    (triangles).}
    \label{fig:privacy_emotion}
\end{figure}

\begin{figure*}[t]
    \centering
    \includegraphics[width=\linewidth]{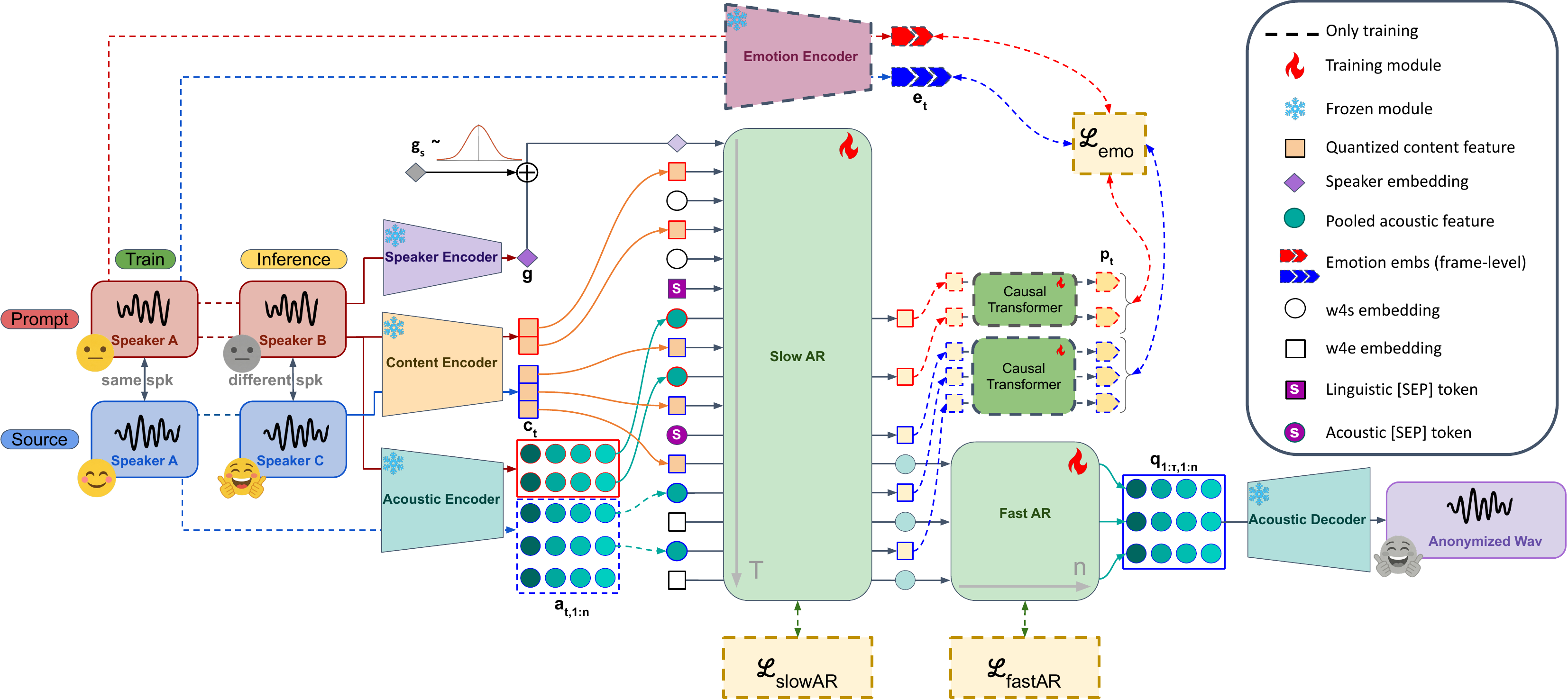}
    \caption{Training and inference configurations.
    \textbf{Training}: prompt and source share the same speaker
    but differ in emotion,
    forcing the model to generate emotional output from source content
    rather than copying prompt-specific patterns.
    Frame-level emotion distillation
    ($\mathcal{L}_{\text{emo}}$, dashed)
    on Slow~AR acoustic hidden states provides additional learning signal.
    \textbf{Inference}: a neutral utterance from the target anonymous speaker
    conceals source identity while the finetuned model
    preserves source emotion;
    no latency is added over the baseline.
    The transformer $f_\theta$ is depicted twice for visual clarity;
    weights are shared.}
    \label{fig:architecture}
\end{figure*}

We propose a supervised finetuning (SFT) approach
with frame-level emotion distillation (Figure~\ref{fig:architecture})
that addresses the root cause of emotion degradation.
Our contributions are:
\begin{itemize}
    \item We show that emotion degradation in NAC-based streaming SA
    is primarily a \emph{training paradigm} problem, not a model capacity issue:
    restructuring training pairs yields 3$\times$ larger gains
    than adding emotional data alone (ablation: Exp1 vs Exp2).
    \item We apply frame-level emotion distillation
    to acoustic token hidden states,
    isolating emotion learning from content supervision (next-token prediction)
    to avoid gradient competition.
    Ablation confirms this design choice:
    acoustic-branch distillation outperforms semantic-branch
    in both UAR and WER (Exp6 vs Exp7).
    \item Our approach achieves the highest reported emotion preservation
    among streaming SA methods (49.2\% UAR, +24\% relative over baseline)
    with \textbf{zero inference overhead}
    and improved privacy (Figure~\ref{fig:privacy_emotion}).
\end{itemize}

\section{Related Work}

\textbf{Speaker Anonymization.}
Traditional SA methods modify speech features
such as pitch and formants~\cite{patino2021mcadams,panariello2024voiceprivacy}
or replace speaker embeddings~\cite{srivastava2020xvector,srivastava2022privacy, kuzmin2024ntunpu, meyer2023gan}.
Recent neural approaches use voice conversion~\cite{meyer2023prosody,yao2025easy}
or NAC-based language
models~\cite{panariello2024nac,kuzmin2026streamvoiceanonenhancingutilityrealtime}.
Most methods prioritize privacy-intelligibility trade-offs,
with emotion preservation receiving less attention~\cite{gaznepoglu2025why}.

\textbf{Streaming SA and Voice Conversion.}
Real-time SA and VC systems face a fundamental trade-off
between latency and quality.
StreamVoice~\cite{wang2024streamvoice} introduced NAC-based language models
for streaming VC with interleaved semantic
and acoustic token generation,
achieving low latency through autoregressive processing.
A key limitation identified in recent work~\cite{tvtsyn2026}
is the temporal mismatch between static speaker embeddings
and dynamic content representations.
While content evolves frame-by-frame,
conventional systems inject speaker identity as a global vector,
leading to over-smoothed timbre and reduced expressivity.
Time-varying timbre (TVT) approaches address this
by allowing speaker characteristics to vary synchronously
with content through attention-based memory mechanisms.
Self-supervised approaches like GenVC~\cite{cai2025genvc}
eliminate dependency on external speaker encoders
through autoregressive generation.
These architectural insights are complementary
to our supervised finetuning approach,
which addresses emotion degradation
from a training paradigm perspective.

\textbf{Emotion in SA.}
Speaker anonymization systems often degrade emotional content~\cite{ghosh2023emostargan,das2023starganvcpp,hua2024emotional},
with the primary cause identified as information loss
in intermediate representations~\cite{gaznepoglu2025why}.
Several approaches address this through
emotion embeddings~\cite{miao2025adapting}
or prosody-aware anonymization~\cite{he2025emotion}.
EASY~\cite{yao2025easy} introduced emotion-aware offline SA
through factorized distillation,
achieving strong emotion preservation (63.8\% UAR)
but requiring full utterance context.
For streaming SA,
StreamVoiceAnon~\cite{kuzmin2026streamvoiceanonenhancingutilityrealtime}
showed limited emotion preservation (39.7\% UAR)
with neutral prompts.
Prior work~\cite{quamer2024slt} achieved low latency
but weak privacy (31.4\% EER),
while DarkStream~\cite{quamer2025darkstream} operates
at \SI{200}{\milli\second} latency
but shows limited emotion preservation (34.7\% UAR).
Recent streaming VC work~\cite{tvtsyn2026} demonstrates
that architectural improvements can improve naturalness,
but these systems still face emotion challenges
where content bottlenecks and speaker replacement
create additional pressure on paralinguistic retention.

\textbf{Knowledge Distillation for Speech.}
Knowledge distillation has been widely applied in speech processing
to transfer representations from large pretrained models
to smaller task-specific models~\cite{chang2022distilhubert}.
Self-supervised models capture emotion-relevant information
in their hidden representations~\cite{pepino2021emotion};
Emotion2Vec+~\cite{ma2024emotion2vec}
provides frame-level emotion representations
through pretraining on large-scale speech data.
Unlike utterance-level emotion labels,
these frame-level features capture fine-grained emotional dynamics
at each time step, making them suitable for streaming applications.
We leverage such features as distillation targets,
carefully selecting the acoustic branch
to avoid interference with content supervision
on the semantic branch.

\section{Proposed Method}

\subsection{Problem Statement}

Streaming anonymization models generally struggle
with emotion preservation due to information bottlenecks
and training objectives that prioritize content
and speaker conversion over paralinguistic attributes.
In NAC-based approaches specifically, we identify two key issues:
(1)  the audio continuation training paradigm encourages the model
 to degrade emotion preservation for the source utterance, and
(2) the discrete token representation loses fine-grained
emotional prosody.

Per-emotion analysis reveals severe imbalance in the baseline:
certain emotions are strongly over-represented
while others are nearly absent
(see Section~\ref{sec:ablation} for details).
This pattern persists even with neutral prompts,
suggesting that the model develops an inherent bias
toward dominant acoustic patterns during pretraining,
likely due to imbalanced emotion distributions in the training data.
Rather than preserving source emotion,
the model defaults to a dominant acoustic style.
The combination of this learned bias and VQ information loss
may create systematic emotion degradation
that simple finetuning on emotional data cannot resolve.

\subsection{Supervised Finetuning with Neutral-Emotion Pairs}

We construct training pairs from an emotional speech corpus
where each pair contains a neutral and an emotional utterance
from the \emph{same speaker}.
This design ensures:
\begin{itemize}
    \item Anonymization still works because speaker embedding mixing
    is unchanged
    \item The model must generate emotional acoustic tokens
    from a neutral prompt
    \item Emotion information must come from source content features,
    not prompt acoustics
\end{itemize}

We also include neutral-to-neutral pairs to maintain balance
and prevent the model from assuming
all source utterances are emotional.

Separate learned separation tokens for the semantic
and acoustic branches mark the boundary
between prompt and source sequences in each token stream.
This explicit boundary helps the model transition
from prompt reproduction to source preservation,
preventing it from copying prompt characteristics
into the source utterance.
These tokens are shown as Linguistic [SEP] and Acoustic [SEP]
in Figure~\ref{fig:architecture}.

\subsection{Frame-Level Emotion Distillation}

We distill frame-level emotion features
from a pretrained emotion extractor into the model's hidden states.
An important design choice is \emph{where} to apply distillation:
\begin{itemize}
    \item \textbf{Semantic branch}: Already supervised
    via next-token prediction (LM loss);
    adding emotion creates gradient competition~\cite{hu2023gradient}
    \item \textbf{Acoustic branch}: No existing supervision;
    clean gradient flow for emotion learning
\end{itemize}

Our architecture (Figure~\ref{fig:architecture})
applies a shared causal transformer $f_\theta$
to acoustic hidden states $h_t^{\text{acou}}$
from the Slow AR branch,
producing predicted emotion embeddings $p_t$,
supervised by $\mathcal{L}_{\text{emo}}$
against Emotion Encoder targets $e_t$:
\begin{equation}
    \mathcal{L}_{\text{emo}} = \frac{1}{T}\sum_{t=1}^{T}
    \lVert p_t - e_t \rVert^2, \quad p_t = f_\theta(h_t^{\text{acou}})
\end{equation}
where $h_t^{\text{acou}}$ are acoustic hidden states,
$e_t$ are frame-level emotion representations
from the Emotion Encoder,
and $p_t$ is the predicted emotion embedding.
While the VQ bottleneck remains at the output stage,
the distillation loss encourages richer emotion encoding
in hidden states prior to quantization.
The total loss combines language
modeling~\cite{kuzmin2026streamvoiceanonenhancingutilityrealtime}
and distillation:
\begin{equation}
    \mathcal{L} = \mathcal{L}_{\text{LM}}
    + w \cdot \mathcal{L}_{\text{emo}}
\end{equation}
where $\mathcal{L}_{\text{LM}} = \mathcal{L}_{\text{slowAR}}
+ \mathcal{L}_{\text{fastAR}}$
are next-token prediction losses
on the Slow AR and Fast AR branches respectively,
and $w$ controls the distillation strength.
Slow AR generates one token per time step along the time axis,
while Fast AR autoregressively generates $n$ codebook tokens
($q_1 \ldots q_n$) per time step using shared weights.
At inference, $f_\theta$ and the Emotion Encoder are removed;
the model operates with the same architecture
and latency as the baseline.

\begin{table}[t]
    \caption{Comparison with prior methods.
    $\uparrow$/$\downarrow$: higher/lower is better.
    Lat.: algorithmic latency in ms.
    EER-L/EER-S: lazy-informed/semi-informed attacker.
    Semi: semi-streaming.
    \textbf{Bold}: best streaming result;
    \underline{underline}: second best.
    Only methods with viable privacy
    (EER-L $\geq$ 40\%) are included,
    following VoicePrivacy evaluation criteria.
    Our two variants correspond to Exp4 (pool-distill) and
    Exp7 (frame-distill) in the ablation study
    (Table~\ref{tab:ablation}).}
    \label{tab:main}
    \centering
    \resizebox{\columnwidth}{!}{%
    \scriptsize
    \begin{tabular}{@{}lcrrrrr@{}}
        \toprule
        \textbf{Method} & \textbf{Type} & \textbf{Lat.} & \textbf{WER}$\downarrow$ & \textbf{UAR}$\uparrow$ & \textbf{EER-L}$\uparrow$ & \textbf{EER-S}$\uparrow$ \\
        \midrule
        Original & -- & -- & 1.83 & 70.07 & 5.16 & -- \\
        \midrule
        EASY~\cite{yao2025easy} & Offline & -- & 2.70 & 63.81 & -- & 45.89 \\
        \midrule
        GenVC-small~\cite{cai2025genvc} & Semi & -- & 8.20 & 34.23 & 48.48 & 15.94 \\
        \midrule
        DarkStream~\cite{quamer2025darkstream} & Online & 200 & 8.75 & 34.73 & 47.26 & \textbf{21.83} \\
        TVTSyn~\cite{tvtsyn2026} & Online & \textbf{80} & 5.35 & 37.32 & 47.55 & 14.57 \\
        StreamVoiceAnon~\cite{kuzmin2026streamvoiceanonenhancingutilityrealtime} &  & &  &  & &  \\
        \quad vctk-1fix & Online & \underline{180} & \textbf{4.54} & 39.72 & 47.19 & 15.92 \\
        \quad crema-emo-4rnd & Online & \underline{180} & 6.59 & 44.59 & 46.53 & \underline{18.63} \\
        \textbf{Ours} &  & &  & &  & \\
        \quad pool-distill & Online & \underline{180} & \underline{5.08} & \underline{46.30} & \underline{48.62} & 18.32 \\
        \quad frame-distill & Online & \underline{180} & 5.77 & \textbf{49.22} & \textbf{48.98} & 18.30 \\
        \bottomrule
    \end{tabular}}
\end{table}

\begin{table*}[t]
    \caption{Ablation study. \cmark~indicates active components.
    StatPool/Causal: aggregation approach.
    Distill: distillation target branch
    (Sem = semantic, Aco = acoustic; -- = acoustic by default).
    \textbf{Bold}: best among Exp1--7;
    \underline{underline}: second best.
    All metrics on IEMOCAP following VoicePrivacy 2024 protocol.}
    \label{tab:ablation}
    \centering
    \setlength{\tabcolsep}{3.5pt}
    \begin{tabular}{@{}lcccccccccccccc@{}}
        \toprule
        & \multicolumn{6}{c}{\textbf{Components}} & \textbf{Content $\downarrow$} & \multicolumn{5}{c}{\textbf{Emotion $\uparrow$}} & \multicolumn{2}{c}{\textbf{Privacy $\uparrow$}} \\
        \cmidrule(lr){2-7} \cmidrule(lr){8-8} \cmidrule(lr){9-13} \cmidrule(lr){14-15}
        \textbf{Model} & \rotatebox{0}{\footnotesize\textbf{FT-CREMA}} & \rotatebox{0}{\footnotesize\textbf{Neu-Emo}} & \rotatebox{0}{\footnotesize\textbf{[SEP]}} & \rotatebox{0}{\footnotesize\textbf{StatPool}} & \rotatebox{0}{\footnotesize\textbf{Causal}} & \rotatebox{0}{\footnotesize\textbf{Distill}} & \textbf{WER} & \textbf{Average} & \textbf{Ang} & \textbf{Hap} & \textbf{Neu} & \textbf{Sad} & \textbf{EER-L} & \textbf{EER-S} \\
        \midrule
        Baseline & & & & & & -- & 4.54 & 39.7 & 35.8 & 81.9 & 33.1 & 8.0 & 47.19 & 15.92 \\
        \midrule
        Exp1 & \cmark & & & & & -- & \textbf{5.00} & 41.1 & 36.3 & \textbf{79.6} & 35.5 & 13.2 & 45.70 & 14.88 \\
        Exp2 & \cmark & \cmark & & & & -- & 5.16 & 45.3 & 35.3 & \underline{75.9} & 48.2 & 21.7 & 47.31 & 16.73 \\
        Exp3 & \cmark & \cmark & \cmark & & & -- & 5.25 & 47.4 & 34.8 & 72.9 & 50.7 & 31.2 & 47.46 & 16.53 \\
        Exp4 & \cmark & \cmark & \cmark & \cmark & & -- & \underline{5.08} & 46.3 & \underline{40.9} & 75.1 & 44.2 & 25.0 & \underline{48.62} & \textbf{18.32} \\
        Exp5 & \cmark & \cmark & \cmark & & \cmark & -- & 5.32 & \underline{48.5} & 40.3 & 65.2 & \textbf{53.6} & \underline{34.8} & 48.19 & 16.78 \\
        Exp6 & \cmark & \cmark & \cmark & & \cmark & Sem & 6.23 & 48.2 & \textbf{48.7} & 66.7 & 49.7 & 27.7 & 47.93 & 17.10 \\
        Exp7 & \cmark & \cmark & \cmark & & \cmark & Aco & 5.77 & \textbf{49.2} & 38.8 & 62.8 & \underline{52.7} & \textbf{42.6} & \textbf{48.98} & \underline{18.30} \\
        \bottomrule
    \end{tabular}
\end{table*}
\section{Experiments}

\subsection{Setup}

We follow the VoicePrivacy 2024 Challenge
protocol~\cite{tomashenko2026vpc3}:
\begin{itemize}
    \item \textbf{Privacy (EER)}: Equal error rate
    from ECAPA-TDNN~\cite{desplanques2020ecapatdnn};
    lazy-informed (trained on original)
    and semi-informed (trained on anonymized) attackers.
    Higher EER = better privacy.
    \item \textbf{Intelligibility (WER)}: Word error rate
    from ASR trained on LibriSpeech. Lower = better.
    \item \textbf{Emotion (UAR)}: Unweighted average recall
    from a speech emotion recognition (SER) model
    trained on IEMOCAP~\cite{busso2008iemocap}
    (4 classes: angry, happy, neutral, sad). Higher = better.
\end{itemize}

We finetune the pretrained open-source
StreamVoiceAnon~\cite{kuzmin2026streamvoiceanonenhancingutilityrealtime}
model (vctk-1fix anonymization strategy, i.e., single fixed target speaker)
on CREMA-D~\cite{cao2014cremad} neutral-emotion pairs
for 5 epochs using 4$\times$ NVIDIA RTX 4090 GPUs
with a learning rate of $1\times10^{-4}$.
Only the Slow~AR and Fast~AR modules are finetuned,
along with the distillation transformer $f_\theta$
(initialized from scratch);
all other components remain frozen.
The [SEP] tokens are randomly initialized learnable embeddings.
We filter CREMA-D to four emotions
(angry, happy, neutral, sad),
excluding fearful and disgusted categories
as these emotions are not present in the IEMOCAP evaluation set.
The dataset provides 7,442 clips from 91 actors,
from which we construct approximately 25,000 neutral-emotion pairs
using many-to-many matching
(all neutral utterances per speaker paired
with all emotional and neutral utterances from the same speaker)
after filtering (quality threshold $q=0.5$
based on vote scores).
We use Emotion2Vec+ large~\cite{ma2024emotion2vec}
(last-layer hidden representations)
as the emotion feature extractor for distillation,
with weight $w=0.01$;
higher values degraded intelligibility
while lower values showed diminishing emotion gains;
training loss plateaued between epochs 5 and 10.
Note that the VoicePrivacy 2024 SER evaluator
is a wav2vec~2.0-based model~\cite{pepino2021emotion}
trained on IEMOCAP via SpeechBrain,
while the distillation teacher (Emotion2Vec+)
uses a data2vec~2.0 backbone,
reducing the risk of circular evaluation.
This evaluator is part of the standard VoicePrivacy protocol
and enables direct comparison with prior work.
The causal transformer $f_\theta$ consists of 2 layers
with hidden dimension matching the acoustic hidden states.
Emotion evaluation uses IEMOCAP development and test sets.

\subsection{Main Results}

Table~\ref{tab:main} presents the results.
Our method's primary achievement is emotion preservation:
49.2\% UAR represents the highest among all streaming methods
(the SER model achieves 70.07\% on original speech,
establishing an upper bound),
while maintaining competitive intelligibility (5.77\% WER)
and strong privacy (48.98\% EER-lazy).

Compared to the StreamVoiceAnon baseline
with neutral prompts (39.7\% UAR),
our method achieves +24\% relative UAR improvement
with a modest WER increase (4.54\% $\to$ 5.77\%)
and improved privacy (EER-lazy: 47.19\% $\to$ 48.98\%).
We hypothesize that emotion distillation
encourages disentanglement of emotional
and speaker information in acoustic hidden states,
reducing identity leakage.
Against the emotion-prompt variant (SVA+EMO, 44.6\% UAR),
we achieve +10\% relative UAR
with better WER (6.59\% $\to$ 5.77\%) and privacy.
Notably, our method uses only a single neutral prompt
instead of four emotion-diverse prompts.

Compared to other streaming methods with viable privacy
(EER-lazy $>$40\%),
our method improves UAR by +32--44\% relative
over GenVC-small (34.2\%), DarkStream (34.7\%),
and TVTSyn (37.3\%),
with comparable or better WER and privacy,
demonstrating that the training paradigm approach
complements architectural improvements.

\subsection{Ablation Study}
\label{sec:ablation}

Table~\ref{tab:ablation} systematically evaluates each component.
All experiments from Exp1 onward use finetuning on CREMA-D;
Exp2 onwards add neutral-emotion pairs;
Exp3 onwards add the separation token;
Exp4--7 explore different distillation architectures.

\textbf{Finetuning on emotional data (Exp1)}:
Simply finetuning on CREMA-D yields only +1.4 UAR,
confirming this is not a data domain issue
but a training paradigm problem.

\textbf{Neutral-emotion pairs (Exp2)}:
Forcing the model to generate emotional output from neutral prompts
provides the largest single improvement (+4.1 UAR),
partially validating our hypothesis
about prompt-driven emotion degradation.
Sad emotion improves from 8.0\% to 21.7\%.

\textbf{Separation token (Exp3)}:
Explicit boundary marking between prompt and source sequences
adds +2.1 UAR, further improving sad to 31.2\%.

\textbf{Emotion distillation architectures (Exp4-5)}:
We explore two aggregation strategies for the distillation head.
Statistical pooling (Exp4) computes mean and standard deviation
over hidden states before prediction,
achieving 46.3\% UAR with 5.08\% WER.
However, Exp4 actually decreases UAR compared to Exp3 (47.4\%),
suggesting that utterance-level pooling
loses temporal emotion dynamics.
A causal transformer (Exp5) maintains frame-level predictions,
achieving 48.5\% UAR with 5.32\% WER,
confirming that temporal modeling
better captures frame-level emotion dynamics.

\textbf{Distillation target selection (Exp6 vs Exp7)}:
Distilling to semantic hidden states (Exp6)
achieves 48.2\% UAR but degrades WER to 6.23\%
due to gradient competition with content supervision.
Acoustic distillation (Exp7) achieves higher UAR (49.2\%)
with better WER (5.77\%),
confirming that the acoustic branch
provides cleaner gradient flow.

\textbf{Per-emotion analysis}:
The most dramatic improvement occurs for ``sad'':
from 8.0\% (baseline) to 42.6\% (ours).
Neutral emotion improves from 33.1\% to 52.7\% (+59\%).
The decrease in ``happy'' (81.9\% $\to$ 62.8\%)
represents correction of over-prediction bias,
as the baseline defaulted to happy-sounding output
regardless of source emotion.

\section{Conclusion}

We have presented a SFT approach
for emotion-preserving streaming speaker anonymization.
Our ablation reveals that emotion preservation
is largely a \emph{training paradigm} problem
for this architecture:
restructuring training pairs
(Exp1$\to$Exp2, +4.2 UAR)
yields 3$\times$ larger gains
than adding emotional data alone
(Baseline$\to$Exp1, +1.4 UAR),
indicating the model has
the capacity to encode emotion but lacks
the right training signal.
Combined with frame-level emotion distillation
on acoustic hidden states,
our method improves the baseline
from 39.7\% to 49.2\% UAR (+24\% relative)
while also improving privacy
(EER-lazy: 47.2\% $\to$ 49.0\%),
with zero inference overhead.

A gap remains compared to offline methods
(EASY: 63.8\% UAR), reflecting
the fundamental latency--quality trade-off:
offline methods access full utterance context,
enabling bidirectional modeling
and utterance-level emotion optimization.
Future work will explore
longer-range emotional context within the causal constraint,
extension to dimensional emotion models,
and adversarial training against adaptive attacks.
While we follow the standard VoicePrivacy 2024 evaluation protocol,
limitations include reliance on a single SER evaluator,
the absence of subjective listening tests,
and evaluation on acted speech corpora only
(both finetuning on CREMA-D and evaluation on IEMOCAP).
Future work should validate on spontaneous emotion corpora
(e.g., MSP-Podcast).

%

\section{Generative AI Use Disclosure}
Generative AI tools were used to assist
with manuscript editing and formatting.
All technical content, experimental design,
and analysis were performed by the authors.

\bibliographystyle{IEEEtran}
\bibliography{references}

@IEEEtranBSTCTL{IEEEexample:BSTcontrol,
  CTLdash_repeated_names = {no}
}

@inproceedings{meyer2025usecases,
  title={Use Cases for Voice Anonymization},
  author={Meyer, Sarina and Vu, Ngoc Thang},
  booktitle={Proc. 5th Symposium on Security and Privacy in Speech Communication (SPSC)},
  year={2025}
}

@inproceedings{patino2021mcadams,
  title={Speaker Anonymisation Using the {McAdams} Coefficient},
  author={Patino, Jose and Tomashenko, Natalia and Todisco, Massimiliano and Nautsch, Andreas and Evans, Nicholas},
  booktitle={Proc. Interspeech 2021},
  pages={1099--1103},
  year={2021},
  doi={10.21437/Interspeech.2021-1069}
}

@inproceedings{srivastava2020xvector,
  title={Design Choices for X-Vector Based Speaker Anonymization},
  author={Srivastava, Brij Mohan Lal and Tomashenko, Natalia and Wang, Xin and Vincent, Emmanuel and Yamagishi, Junichi and Todisco, Massimiliano and Nautsch, Andreas and Shan, Junwon and Evans, Nicholas},
  booktitle={Proc. Interspeech 2020},
  pages={2512--2516},
  year={2020},
  doi={10.21437/Interspeech.2020-2692}
}

@article{tomashenko2022vpc2020,
  title={The {VoicePrivacy} 2020 Challenge: Results and Findings},
  author={Tomashenko, Natalia and Wang, Xin and Vincent, Emmanuel and Patino, Jose and Srivastava, Brij Mohan Lal and No{\'e}, Pierre-Gabriel and Nautsch, Andreas and Evans, Nicholas and Yamagishi, Junichi and O'Brien, Benjamin and Chanclu, Ana{\"i}s and Bonastre, Jean-Fran{\c{c}}ois and Todisco, Massimiliano and Maouche, Mohamed},
  journal={Computer Speech \& Language},
  volume={74},
  pages={101362},
  year={2022},
  doi={10.1016/j.csl.2022.101362}
}

@article{srivastava2022privacy,
  title={Privacy and Utility of X-Vector Based Speaker Anonymization},
  author={Srivastava, Brij Mohan Lal and Maouche, Mohamed and Sahidullah, Md and Vincent, Emmanuel and Bellet, Aur{\'e}lien and Tommasi, Marc and Tomashenko, Natalia and Wang, Xin and Yamagishi, Junichi},
  journal={IEEE/ACM Transactions on Audio, Speech and Language Processing},
  volume={30},
  pages={2383--2395},
  year={2022},
  doi={10.1109/TASLP.2022.3190741}
}

@inproceedings{meyer2023gan,
  title={Anonymizing Speech with Generative Adversarial Networks to Preserve Speaker Privacy},
  author={Meyer, Sarina and Tilli, Pascal and Denisov, Pavel and Lux, Florian and Koch, Julia and Vu, Ngoc Thang},
  booktitle={Proc. IEEE Spoken Language Technology Workshop (SLT)},
  pages={912--919},
  year={2023}
}

@inproceedings{hua2024emotional,
  title={Emotional Speech Anonymization: Preserving Emotion Characteristics in Pseudo-Speaker Speech Generation},
  author={Hua, Hua and Shang, Zengqiang and Li, Xuyuan and Shi, Peiyang and Yang, Chen and Wang, Li and Zhang, Pengyuan},
  booktitle={Proc. 4th Symposium on Security and Privacy in Speech Communication (SPSC)},
  pages={55--60},
  year={2024}
}

@article{tomashenko2026vpc3,
  title={The Third {VoicePrivacy} Challenge: Preserving Emotional Expressiveness and Linguistic Content in Voice Anonymization},
  author={Tomashenko, Natalia and Miao, Xiaoxiao and Champion, Pierre and Meyer, Sarina and Panariello, Michele and Wang, Xin and Evans, Nicholas and Vincent, Emmanuel and Yamagishi, Junichi and Todisco, Massimiliano},
  journal={arXiv preprint arXiv:2601.11846},
  year={2026}
}

@article{panariello2024voiceprivacy,
  title={The {VoicePrivacy} 2022 Challenge: Progress and Perspectives in Voice Anonymisation},
  author={Panariello, Michele and Tomashenko, Natalia and Wang, Xin and Miao, Xiaoxiao and Champion, Pierre and Nourtel, Hubert and Todisco, Massimiliano and Evans, Nicholas and Vincent, Emmanuel and Yamagishi, Junichi},
  journal={IEEE/ACM Transactions on Audio, Speech and Language Processing},
  volume={32},
  pages={3497--3512},
  year={2024},
  publisher={IEEE},
  doi={10.1109/TASLP.2024.3430530}
}

@inproceedings{wang2024streamvoice,
  title={{StreamVoice}: Streamable Context-Aware Language Modeling for Real-time Zero-Shot Voice Conversion},
  author={Wang, Zhichao and Chen, Yuanzhe and Wang, Xinsheng and Xie, Lei and Wang, Yuping},
  booktitle={Proceedings of the 62nd Annual Meeting of the Association for Computational Linguistics (Volume 1: Long Papers)},
  pages={7328--7338},
  year={2024},
  address={Bangkok, Thailand},
  publisher={Association for Computational Linguistics}
}

@inproceedings{kuzmin2026streamvoiceanonenhancingutilityrealtime,
  title={Stream-Voice-Anon: Enhancing Utility of Real-Time Speaker Anonymization via Neural Audio Codec and Language Models},
  author={Kuzmin, Nikita and Liu, Songting and Lee, Kong Aik and Chng, Eng Siong},
  booktitle={Proc. ICASSP 2026},
  year={2026}
}

@inproceedings{yao2025easy,
  title={{EASY}: Emotion-aware Speaker Anonymization via Factorized Distillation},
  author={Yao, Jixun and Liu, Hexin and Chng, Eng Siong and Xie, Lei},
  booktitle={Proc. Interspeech 2025},
  pages={3219--3223},
  year={2025},
  doi={10.21437/Interspeech.2025-XXXX}
}

@inproceedings{quamer2025darkstream,
  title={{DarkStream}: Real-time Speech Anonymization with Low Latency},
  author={Quamer, Waris and Gutierrez-Osuna, Ricardo},
  booktitle={Proc. IEEE Automatic Speech Recognition and Understanding Workshop (ASRU)},
  year={2025}
}

@inproceedings{quamer2024slt,
  title={End-to-end Streaming Model for Low-Latency Speech Anonymization},
  author={Quamer, Waris and Gutierrez-Osuna, Ricardo},
  booktitle={Proc. IEEE Spoken Language Technology Workshop (SLT)},
  pages={},
  year={2024},
  organization={IEEE}
}

@inproceedings{tvtsyn2026,
  title={{TVTSyn}: Content-Synchronous Time-Varying Timbre for Streaming Voice Conversion and Anonymization},
  author={Quamer, Waris and Tseng, Mu-Ruei and Nasrallah, Ghady and Gutierrez-Osuna, Ricardo},
  booktitle={Proc. International Conference on Learning Representations (ICLR)},
  year={2026}
}

@inproceedings{cai2025genvc,
  title={{GenVC}: Self-Supervised Zero-Shot Voice Conversion},
  author={Cai, Zexin and Li, Henry Xinyuan and Garg, Ashi and Garc{\'\i}a-Perera, Leibny Paola and Duh, Kevin and Khudanpur, Sanjeev and Wiesner, Matthew and Andrews, Nicholas},
  booktitle={Proc. IEEE Automatic Speech Recognition and Understanding Workshop (ASRU)},
  year={2025},
  organization={IEEE}
}

@inproceedings{ma2024emotion2vec,
  title={emotion2vec: Self-Supervised Pre-Training for Speech Emotion Representation},
  author={Ma, Ziyang and Zheng, Zhisheng and Ye, Jiaxin and Li, Jinchao and Gao, Zhifu and Zhang, Shiliang and Chen, Xie},
  booktitle={Findings of the Association for Computational Linguistics: ACL 2024},
  pages={15747--15760},
  year={2024},
  address={Bangkok, Thailand},
  publisher={Association for Computational Linguistics}
}

@article{cao2014cremad,
  title={{CREMA-D}: Crowd-Sourced Emotional Multimodal Actors Dataset},
  author={Cao, Houwei and Cooper, David G and Keutmann, Michael K and Gur, Ruben C and Nenkova, Ani and Verma, Ragini},
  journal={IEEE Transactions on Affective Computing},
  volume={5},
  number={4},
  pages={377--390},
  year={2014},
  publisher={IEEE},
  doi={10.1109/TAFFC.2014.2336244}
}

@article{busso2008iemocap,
  title={{IEMOCAP}: Interactive Emotional Dyadic Motion Capture Database},
  author={Busso, Carlos and Bulut, Murtaza and Lee, Chi-Chun and Kazemzadeh, Abe and Mower, Emily and Kim, Samuel and Chang, Jeannette N and Lee, Sungbok and Narayanan, Shrikanth S},
  journal={Language Resources and Evaluation},
  volume={42},
  number={4},
  pages={335--359},
  year={2008},
  publisher={Springer},
  doi={10.1007/s10579-008-9076-6}
}

@inproceedings{desplanques2020ecapatdnn,
  title={{ECAPA-TDNN}: Emphasized Channel Attention, Propagation and Aggregation in {TDNN} Based Speaker Verification},
  author={Desplanques, Brecht and Thienpondt, Jenthe and Demuynck, Kris},
  booktitle={Proc. Interspeech 2020},
  pages={3830--3834},
  year={2020},
  doi={10.21437/Interspeech.2020-2650}
}

@inproceedings{ghosh2023emostargan,
  title={Emo-{StarGAN}: A Semi-Supervised Any-to-Many Non-Parallel Emotion-Preserving Voice Conversion},
  author={Ghosh, Suhita and Das, Arnab and Sinha, Yamini and Siegert, Ingo and Polzehl, Tim and Stober, Sebastian},
  booktitle={Proc. Interspeech 2023},
  pages={2093--2097},
  year={2023},
  doi={10.21437/Interspeech.2023-191}
}

@inproceedings{das2023starganvcpp,
  title={{StarGAN-VC++}: Towards Emotion Preserving Voice Conversion Using Deep Embeddings},
  author={Das, Arnab and Ghosh, Suhita and Polzehl, Tim and Siegert, Ingo and Stober, Sebastian},
  booktitle={Proc. 12th ISCA Speech Synthesis Workshop (SSW)},
  pages={81--87},
  year={2023},
  doi={10.21437/ssw.2023-13}
}

@inproceedings{gaznepoglu2025why,
  title={Why disentanglement-based speaker anonymization systems fail at preserving emotions?},
  author={Gaznepoglu, {\"U}nal Ege and Peters, Nils},
  booktitle={Proc. ICASSP 2025},
  pages={1--5},
  year={2025},
  doi={10.1109/ICASSP49660.2025.10889709}
}

@inproceedings{cai2024privacy,
  title={Privacy Versus Emotion Preservation Trade-Offs in Emotion-Preserving Speaker Anonymization},
  author={Cai, Zexin and Li Xinyuan, Henry and Garg, Ashi and Garc{\'\i}a-Perera, Leibny Paola and Duh, Kevin and Khudanpur, Sanjeev and Andrews, Nicholas and Wiesner, Matthew},
  booktitle={Proc. IEEE Spoken Language Technology Workshop (SLT)},
  pages={409--414},
  year={2024},
  doi={10.1109/SLT61566.2024.10832351}
}

@inproceedings{he2025emotion,
  title={Emotion-Preserving Prosody Anonymization Network for Voice Privacy Protection},
  author={He, Jiabei and Zhao, Shiwan and Zhou, Jiaming and Sun, Haoqin and Wang, Hui and Qin, Yong},
  booktitle={Proc. ICASSP 2025},
  pages={1--5},
  year={2025},
  doi={10.1109/ICASSP49660.2025.10890338}
}

@article{miao2025adapting,
  title={Adapting general disentanglement-based speaker anonymization for enhanced emotion preservation},
  author={Miao, Xiaoxiao and Zhang, Yuxiang and Wang, Xin and Tomashenko, Natalia and Soh, Donny Cheng Lock and Mcloughlin, Ian},
  journal={Computer Speech \& Language},
  volume={94},
  pages={101810},
  year={2025},
  doi={10.1016/j.csl.2025.101810}
}

@inproceedings{zhang2024speechtokenizer,
  title={{SpeechTokenizer}: Unified Speech Tokenizer for Speech Language Models},
  author={Zhang, Xin and Zhang, Dong and Li, Shimin and Zhou, Yaqian and Qiu, Xipeng},
  booktitle={Proc. International Conference on Learning Representations (ICLR)},
  year={2024}
}

@inproceedings{chang2022distilhubert,
  title={{DistilHuBERT}: Speech Representation Learning by Layer-wise Distillation of Hidden-unit {BERT}},
  author={Chang, Heng-Jui and Yang, Shu-wen and Lee, Hung-yi},
  booktitle={Proc. ICASSP 2022},
  pages={7087--7091},
  year={2022},
  doi={10.1109/ICASSP43922.2022.9747490}
}

@inproceedings{hu2023gradient,
  title={Gradient Remedy for Multi-Task Learning in End-to-End Noise-Robust Speech Recognition},
  author={Hu, Yuchen and Chen, Chen and Li, Ruizhe and Zhu, Qiushi and Chng, Eng Siong},
  booktitle={Proc. ICASSP 2023},
  pages={1--5},
  year={2023},
  doi={10.1109/ICASSP49357.2023.10096615}
}

@inproceedings{pepino2021emotion,
  title={Emotion Recognition from Speech Using wav2vec 2.0 Embeddings},
  author={Pepino, Leonardo and Riera, Pablo and Ferrer, Luciana},
  booktitle={Proc. Interspeech 2021},
  pages={3400--3404},
  year={2021},
  doi={10.21437/Interspeech.2021-703}
}

@inproceedings{kuzmin2024ntunpu,
  title     = {{NTU-NPU System for Voice Privacy 2024 Challenge}},
  author    = {Kuzmin, Nikita and Luong, Hieu-Thi and Yao, Jixun and Xie, Lei and Lee, Kong Aik and Chng, Eng-Siong},
  year      = {2024},
  booktitle = {Proc. 4th Symposium on Security and Privacy in Speech Communication (SPSC)},
  pages     = {72--79},
  doi       = {10.21437/SPSC.2024-13},
}

@inproceedings{yao2024npuntu,
  title     = {{NPU-NTU System for Voice Privacy 2024 Challenge}},
  author    = {Jixun Yao and Nikita Kuzmin and Qing Wang and Pengcheng Guo and Ziqian Ning and Dake Guo and Kong Aik Lee and Eng-Siong Chng and Lei Xie},
  year      = {2024},
  booktitle = {{4th Symposium on Security and Privacy in Speech Communication}},
  pages     = {67--71},
  doi       = {10.21437/SPSC.2024-12},
}

@inproceedings{panariello2024nac,
  title={Speaker Anonymization Using Neural Audio Codec Language Models},
  author={Panariello, Michele and Nespoli, Francesco and Todisco, Massimiliano and Evans, Nicholas},
  booktitle={Proc. ICASSP 2024},
  pages={4725--4729},
  year={2024},
  doi={10.1109/ICASSP48485.2024.10447871}
}

@inproceedings{meyer2023prosody,
  title={Prosody Is Not Identity: A Speaker Anonymization Approach Using Prosody Cloning},
  author={Meyer, Sarina and Lux, Florian and Koch, Julia and Denisov, Pavel and Tilli, Pascal and Vu, Ngoc Thang},
  booktitle={Proc. ICASSP 2023},
  pages={1--5},
  year={2023},
  doi={10.1109/ICASSP49357.2023.10096607}
}

\end{document}